\documentclass[aps,showpacs,nofootinbib,12pt]{revtex4-2}
\usepackage{graphicx}
\usepackage{amssymb}
\usepackage{amsmath}
\usepackage{graphics}
\usepackage{dcolumn}
\usepackage{bm}
\usepackage{color}
\usepackage{epstopdf}
\usepackage{lipsum}
\usepackage{hyperref} 

\begin{document}

	\newcommand{\blue}[1]{\textcolor{blue}{#1}}
	\newcommand{\new}{\blue}
	\newcommand{\green}[1]{\textcolor{green}{#1}}
	\newcommand{\modif}{\green}
	\newcommand{\red}[1]{\textcolor{red}{#1}}
	\newcommand{\attention}{\red}
	
	%
	%
	
	\title{Chiral Symmetry Restoration using the Running Coupling Constant from the Light-Front Approach to QCD}

	\author{S. D.  Campos\footnote{sergiodc@ufscar.br}}

	\address{Applied Mathematics Laboratory-CCTS/DFQM, Federal University of S\~ao Carlos,\\ Sorocaba CEP 18052780, Brazil}
	
	
	
	\begin{abstract}
	In this work, the distance between a quark-antiquark pair is analyzed through both the confinement potential as well as the hadronic total cross section.  Using the Helmholtz free energy, entropy is calculated near the minimum of the total cross section through the confinement potential. A fitting procedure for the proton-proton total cross section is performed, defining the fitting parameters. Therefore, the only free parameter remaining in the model is the mass scale $\kappa$ used to define the running coupling constant of the light-front approach to QCD. The mass scale controls the distance $r$ between the quark-antiquark pair and, under some conditions, it allows the occurrence of free quarks even in the confinement regime of QCD.
		
	\end{abstract}
{}











\maketitle
\section{Introduction}
Chiral symmetry restoration is one of the most interesting issues in the phase diagram of Quantum Chromodynamics (QCD). As well-known, there is evidence of a phase transition from confinement to non-confinement phase at the saturation scale and predicted, for example, by lattice simulations \cite{Y.Aoki.S.Borsanyi.S.Durr.Z.Fodor.S.D.Katz.S.Krieg.K.K.Szabo.JHEP.0906.088.2009,S.Borsanyi.etal.Wuppertal.Budapest.Collaboration.JHEP.1009.073.2010,C.Ratti.Rept.Prog.Phys.81.8.084301.2018,A.Bazavov.etal.USQCD.Collaboration.Eur.Phys.J.A55.11.194.2019}. These results were supported by recent analyses performed by the Beam Energy Scan (BES) program \cite{L.Adamczyk.etal.STAR.Collaboration.Phys.Rev.C.96.4.044904.2017,A.Andronic.P.Braun-Munzinger.K.Redlich.J.Stachel.Nature.561.7723.321.2018} on experimental data from Relativistic Heavy-Ion Collisions. 

As a fundamental ingredient, the internal entropy of colliding hadrons should be taken into account to offer a complete view of the chiral symmetry restoration in QCD. However, the proper definition of entropy is a hard task in any physical context. For example, entropy can be related to one-quarter of the black hole event horizon area 
or Shannon's information disorder approach. 
Moreover, the correct definition of entropy is a problem that permeates several issues as, for example, the study of entropy production in rotating black holes \cite{cvetic}, in the so-called Ekpyrotic universe \cite{li}, in Szekeres spacetimes \cite{herrera}, in quark-gluon plasma \cite{mattiello}, in light clusters \cite{vermani}, and decoherence processes \cite{fries}, among others. 

Recently, a maximal entropy calculation was introduced assuming saturation effects due to the properties of the factorizable gluon density \cite{K.Kutak.Phys.Lett.B705.217.2011}. This result has closely connected with those obtained previously by Kharzeev and Tuchin \cite{tuchin}, whose achievements are due to the use of the Unruh effect 
(analogous to the Hawking effect) 
applied to hadrons in the accelerated rest frame. 
They found a transition temperature $T_s\approx Q_s/2\pi$ at the spontaneous symmetry breaking, where $Q_s$ is the transferred momentum in the gluon rest frame. 

On the other hand, a possible topological phase transition in the hadron was proposed in Ref. \cite{S.D.Campos.Arxiv.2020}, allowing a chiral symmetry breaking below the saturation scale with a transition temperature $T_c\approx 0.001$ GeV, far below the Hagedorn temperature. Thus, above $T_c$ but below the Hagedorn temperature \cite{R.Hagedorn.NuovoCim.Suppl.3.147.1965,R.Hagedorn.NuovoCim.56A.1027.1968}, the appearance of a free quark state is allowed, resulting in an effect quite similar to the emergence of the free vortex in the XY model. It should be emphasized that was predicted previously  a chiral symmetry breaking by McLerran and Pisarski \cite{L.McLerran.R.D.Pisarski.Nucl.Phys.A796.83.2007} as well by Glozman and Wagenbrunn \cite{L.YA.Glozman.R.F.Wagenbrunn.Mod.Phys.Lett.A23.2385.2008}, whose possible explanation is given by the presence of quarkyonic matter, introduced years ago \cite{L.McLerran.R.D.Pisarski.Nucl.Phys.A796.83.2007,L.YA.Glozman.R.F.Wagenbrunn.Mod.Phys.Lett.A23.2385.2008}. In this scenario, the chiral symmetry is restored even in the presence of hadronic matter.  


In the present work, one supposes that the hadronic total cross section, $\sigma_{tot}(s)$, can be divided into a finite number of non-interacting disjoint 2D cells, each one containing a quark-antiquark ($q\bar{q}$) pair ($u\bar{u}$ and $d\bar{d}$) separated by a distance $r(s)$, depending on the energy $\sqrt{s}$ measured in the center-of-mass system \cite{S.D.Campos.Arxiv.2020}. This approach is analogous to the so-called Berezinskii-Kosterlitz-Thouless (BKT) phase transition, which occurs without spontaneous symmetry breaking \cite{V.L.Berezinskii.Zh.Eksp.Teor.Fiz.59.907.1970,J.M.Kosterlitz.D.J.Thouless.J.Phys.C6.1181.1973}. In that planar model, at low temperatures, the presence of nontrivial vortex configuration is suppressed due to the binding effect caused by the vortex-antivortex interaction. At high temperatures above a critical value, the binding effect can be neglected, and the nontrivial vortex state emerges. 

The main goal of the present work is to take into account the physical consequences of the confinement potential inside the cell containing the $q\bar{q}$-pair considering a naive model with finite temperature and null quark chemical potential. Then, the free quark state is achieved when the spatial separation, $r(s)$, between quark and antiquark is less than $r_0$, the confining scale. Using some constraints, one obtains an asymptotic bound to the rise of $r(s)$ as the collision energy grows. Moreover, using the running coupling from the light-front holographic QCD (see \cite{A.Deur.S.J.Brodsky.G.F.deTeramond.2016} and references therein), the non-confinement regime is achieved below the saturation scale for $\kappa\approx 0.002$ GeV. The mass scale $\kappa$ emerges in de Alfaro, Fubini, and Furlan's attempt to build a fully invariant conformal Lagrangian theory \cite{V.deAlfaro.S.Fubini.G.Furlan.Il.Nuovo.Cimento.34.4.569.1976}. In the example they gave in the original work, the mass scale provides the necessary infrared cut-off of the theory, and once determined the scale furnishes the rest of the unknown parameters of the problem. In the present context presented here, $\kappa$ is used as a free parameter representing the $e\!f\!f\!ective$ mass of the cell, being straightly connected with the distance $r(s)$ as shall be seen.  

The manuscript is organized as follows. In section \ref{app}, one introduces the relation for the entropy, total cross section, and the distance between the quark and antiquark in the cell. In section \ref{zerocp}, one shows that, under some conditions, the possible existence of a chiral symmetry restoration in the confinement phase of QCD. Final remarks are left for section \ref{fr}.


\section{\label{app}Total Cross Section and Entropy}


First of all, it should be stressed that due to the Fitzgerald-Lorentz length contraction, the hadron contracts in the direction of its motion, although its visual appearance, which depends exclusively on its motion, is invariant in special relativity \cite{J.Terrell.Phys.Rev.116.4.1041.1959,R.Penrose.Math.Proc.Cambridge.Philos.Soc.55.137.1959}. Thus, a relativistically moving sphere is not an ellipsoid, but it would be a rotated sphere \cite{R.Penrose.Math.Proc.Cambridge.Philos.Soc.55.137.1959}. However, as the speed approaches $c$ in its proper reference frame, the hadron tends to become a thin disk. Then, it is reasonable to suppose that some physical properties observed in 2D planar objects may occur in the hadron \cite{S.D.Campos.Arxiv.2020}. Here, we maintain the analogy between the hadron in high-energy collisions and 2D planar objects as done elsewhere \cite{S.D.Campos.Arxiv.2020}. Thus, one assumes that hadronic total cross section can be divided into a finite number of disjoint cells containing a single $q\bar{q}$-pair, that interacts with each other through a confinement potential. Then, the entropy is written as \cite{S.D.Campos.Arxiv.2020}
\begin{eqnarray}\label{eq:sdc_1}
	S(s)=\ln \left(\frac{\sigma_{tot}(s)}{\pi r^2}\right),
\end{eqnarray}

\noindent where $r=r(s)$ is the radius of the cell containing the $q\bar{q}$-pair. The above assumption was based on the well-known behavior of 2D planar objects with vortex-antivortex pairs, where the presence of a BKT phase transition allows the emergence of new degrees of freedom and the consequent emergence of a free vortex \cite{V.L.Berezinskii.Zh.Eksp.Teor.Fiz.59.907.1970,J.M.Kosterlitz.D.J.Thouless.J.Phys.C6.1181.1973}. 

Although the experimental access to $r(s)$ is forbidden at the present-day energies, the theoretical determination of some physical bounds is possible, in principle, by using high-energy theorems as well as taking into account different approaches to entropy $S(s)$. Therefore, one can rewrite relation (\ref{eq:sdc_1}) as
\begin{eqnarray}\label{eq:sdc_2}
	r^2=\frac{1}{\pi}\sigma_{tot}(s)e^{-S(s)}.
\end{eqnarray}


A remarkable realization of Axiomatic Quantum Field Theory in the early 1960s is the Froissart-Martin bound. 
However, despite its beauty, the Froissart-Martin result is an asymptotic inequality: It does not give us a hint about the behavior of $\sigma_{tot}(s)$ at intermediate energies or even the "correct" functional form of $\sigma_{tot}(s)$ at high energies. That bound is usually written as
\begin{eqnarray}\label{froissart} 
	\sigma_{tot}(s)\leq \frac{\pi}{m_{\pi}^2}\ln^2(s/s_0),
\end{eqnarray}

\noindent where $m_{\pi}$ is pion mass and $\sqrt{s_0}$ is some initial energy (usually $\sqrt{s_0}=1.0$ GeV). Of course, for $0<e^{-S(s)}$, one can write the inequality
\begin{eqnarray}\label{eq:sdc_bound}
	r^2\leq \left[\frac{1}{m_{\pi}^2}\ln^2(s/s_0)\right]e^{-S(s)},
\end{eqnarray}

\noindent which acts as an upper bound for $r$. Notice, in addition, that the total cross section is related to the imaginary part of the forward scattering amplitude through the well-known optical theorem 
\begin{eqnarray}
	\sigma_{tot}(s)=\frac{\mathrm{Im}F(s)}{s},
\end{eqnarray}

\noindent where $F(s)$ is the forward scattering amplitude. Thus, one can also write
\begin{eqnarray}\label{eq:opt_theo}
	r^2=\left[\frac{\mathrm{Im}F(s)}{s}\right]e^{-S(s)}.
\end{eqnarray}

The usage of specific parameterizations for $\sigma_{tot}(s)$ can help to understand the behavior of $r$ depending on $\sqrt{s}$. In addition, some known results can also be used: For example, for small values of transferred momentum $\sqrt{|t|}$ in the center-of-mass system, the scattering amplitude is mainly given by its absorptive part. Then, one can use the differential cross section, for example, for $\sqrt{|t|}<0.1$ GeV, to infer effects introduced by $t$ in Eq. (\ref{eq:opt_theo}). 

On the other hand, considering the result given by Eq. (\ref{eq:sdc_bound}), we can analyze the distance $r$ according to $e^{-S(s)}$. 

Firstly, consider the case where $S(s)$ is not a logarithmic function of its variable. Then for an increasing entropy of the form $S(s)\sim s/s_0$ (for example), the distance $r$ fastly decreases, leading to the presence of free quarks due to chiral symmetry restoration. In contrast, for a system with decreasing entropy of the form $S(s)\sim s_0/s$, the distance $r$ increases as the collision energy grow. 

Secondly, considering that $S(s)$ is a logarithmic function of its variable. Then for a system with a slowing increasing entropy given by $S(s)\sim \ln(s/s_0)$, $r$ tends to zero for sufficiently high energy. Of course, the decreasing entropy case is given by $S(s)\sim \ln(s_0/s)$, which implies the increase of $r$ as $s$ grow. Both situations where entropy decreases as the collision energy grows, although interesting, are not treated here.


It is important to stress that the definition of $S(s)$ only makes sense here if one takes into account the relative distance between the $q$ and $\bar{q}$. Thus, $S(s)$ should explicitly depend on $r$. Following the approach presented here, one introduces the distance $r$ in $S(s)$ considering the confinement potential, as shall be seen.

\section{\label{zerocp}Finite Temperature: Far Below the Saturation Scale}

\subsection{Confinement Potential and Running Coupling in the Confinement Phase of QCD}

As a key point for the problem, the internal energy of the colliding hadron cannot be deduced either from the first principles of QCD or thermodynamics. Thus the use of potential energy may be a useful approach to get some insights into the physical processes, as performed since the 1950s \cite{D.Bohm.Phys.Rev.85.166.1952.ibid.180.1952,Sh.F.Y.Liu_R.Rapp.arXiv:1501.07892.2015,G.Dennis.M.A.de.Gosson.B.J.Hiley.Phys.Lett.A378.2363.2014,G.Dennis.M.A.de.Gosson.B.J.Hiley.Phys.Lett.A379.1224.2015}. 

There are several confinement potentials in the literature (for example, see \cite{C.Quigg.J.L.Rosner.Phys.Lett.B71.153.1977,C.Quigg.J.L.Rosner.Phys.Rep.56(4).167.1979,e_eichten_Phys_Rev_Lett_34_369_1975,e_eichten_Phys.Rev.D17.3090.1978,e_eichten_Phys.Rev.D21.203.1980}). For the sake of simplicity, one considers only the Cornell confinement potential \cite{e_eichten_Phys_Rev_Lett_34_369_1975,e_eichten_Phys.Rev.D17.3090.1978,e_eichten_Phys.Rev.D21.203.1980}, which presents a good agreement with experimental data for the spectrum of light and heavy mesons \cite{M.G.Olsson.S.Vesell.K.Williams.Phys.Rev.D51.5079.1995,D.Ebert.V.O.Galkin.R.N.Faustov.Phys.Rev.D57.5663.1998,E.J. Eichten.C.Quigg.Phys.Rev.D49.5845.1994}. That potential is usually written as
\begin{eqnarray}\label{eq:conf_pot}
	V(r)=-\frac{4}{3}\frac{\alpha_s(r)}{r}+\sigma r,
\end{eqnarray}

\noindent where $\alpha_s(r)$ is the Fourier sine transform of the running coupling constant $\alpha_s(Q)$, and $\sqrt{\sigma}$ is the string tension. For static-source $\sqrt{\sigma}=0.46$ GeV \cite{S.Aoki.etal.PRD79.034503.2009} while an average estimation $\sqrt{\sigma}=0.405$ GeV is obtained for cold strongly interacting matter \cite{PRD-90-074017-2014}. Without loss of generality, one adopts hereafter $\sqrt{\sigma}=0.4$ GeV. 

The running coupling constant of QCD is far from a unified definition in the confinement regime of QCD. Although the perturbative sector of QCD has a well-defined definition, the same does not occur with the non-perturbative regime, where several approaches can be used. Two basic examples of definitions are the "analytic" running coupling obtained by Shirkov and Solovtsov by using the spectral density from K\"all\'en-Lehmann relation \cite{D.V.Shirkov.I.L.Solovtsov.Phys.Rev.Lett.79.1209.1997} and the running coupling from the light-front approach to QCD \cite{S.J.Brodsky.G.FdeTeramond.H.G.Dosch.C.Lorce.IJMPA31.19.1630029.2016}. 

In this work, one uses the light-front holographic QCD approach to the running coupling \cite{S.J.Brodsky.G.F.deTeramond.A.Deur.Phys.Rev.D81.096010.2010,A.Deur.S.J.Brodsky.G.F.deTeramond.2016}, whose choice is based on both its simplicity as well good description of the available experimental data  \cite{A.Deur.S.J.Brodsky.G.F.deTeramond.2016}. In the light-front holographic approach, the running coupling is set for all values of $Q$ as \cite{A.Deur.S.J.Brodsky.G.F.deTeramond.2016}
\begin{eqnarray}\label{eq:bdtd}
	\alpha_s(Q)=e^{-Q^2/4 \kappa^2},
\end{eqnarray}

\noindent where $\kappa$ is the mass scale (also called mass parameter) determined from the soft-wall model \cite{H.G.Dosch.J.Erlich.Phys.Rept.584.1.2015,A.Karch.E.Katz.D.T.Son.M.A.Stephanov.PhysRevD.74.015005.2006}, and defined for the hadron as
\begin{eqnarray}\label{eq:k_definition}
	M^2=4\kappa^2(L+S/2+n),
\end{eqnarray}

\noindent where $M$ is the hadron mass, $L$, $S$, and $n$ refer to internal orbital angular momentum, internal spin, and radial quantum number, respectively. The soft-wall model is based on a quadratic dilaton field \cite{A.Karch.E.Katz.D.T.Son.M.A.Stephanov.PhysRevD.74.015005.2006} that leads to a harmonic confining potential whose strength is measured by $\kappa$. The consequence of such potential is the breaking of conformal invariance. 

It is important to stress that $\kappa$ has a value that depends on the meson Regge trajectories analyzed, and settle down, in general, in the range $0.3 \lesssim \kappa \lesssim 0.8$ GeV \cite{H.G.Dosch.J.Erlich.Phys.Rept.584.1.2015}. Notice, in addition, that the linear Regge trajectories given by (\ref{eq:k_definition}) were obtained due to the quadratic dilaton field approach \cite{A.Karch.E.Katz.D.T.Son.M.A.Stephanov.PhysRevD.74.015005.2006}. However, this model entangles the explicit and spontaneous breaking of chiral symmetries \cite{P.Zhang.J.High.Energ.Phys.2010.5.39.2010}. It is also important to stress that the linearity of Regge trajectories is not true everywhere, being more evident for light baryons and mesons \cite{A.J.G.Hey.R.L.Kelly.Phys.Rep.96.71.1983}. Thus, definition (\ref{eq:k_definition}) may not hold everywhere. 

Generally, is necessary a value of $\kappa\approx 0.5$ GeV to describe mesons and baryon masses as well as the assumption of linear Regge trajectories \cite{S.J.Brodsky.G.FdeTeramond.H.G.Dosch.C.Lorce.IJMPA31.19.1630029.2016,T.Branz.T.Gutsche.V.L.Lyubovitskij.I.Schmidt.A.Vega.Phys.Rev.D82.074022.2010}. This value also satisfactorily describes the mesons trajectories of the $\rho$ and $K^*$ \cite{H.G.Dosch.J.Erlich.Phys.Rept.584.1.2015}. In contrast, from fitting the nucleon form factors in the Anti-de Sitter (AdS)/QCD soft-wall model, one has $\kappa\approx 0.40$ GeV \cite{D.Chakrabarti.C.Mondal.Eur.Phys.JC.73.2671.2013}. Moreover, from form factors analysis, the values of $\kappa$ are lower than those for Regge trajectories and mass spectrum \cite{H.G.Dosch.J.Erlich.Phys.Rept.584.1.2015}. For example, $\kappa=0.261 \pm 0.002$ GeV assuming a constant current quark mass $m_q=0.005$ GeV in a model inspired in the AdS/QCD correspondence \cite{A.Bacchetta.S.Cotogno.B.Pasquini.Phys.Lett.B771.546.2017}. 

As aforementioned, one supposes that mass scale $\kappa$ represents the $e\!f\!f\!ective$ mass of the $q\bar{q}$-pair in the cell and, then, one assumes $\kappa<\!<1$ GeV. This constraint will be clarified later in this work.




Returning to the running coupling constant. To avoid the use of the mnemonic rule from Quantum Mechanics $Q\sim 1/r$, one should use the Fourier sine transform of Eq. (\ref{eq:bdtd}) since, as noted by Shirkov, this rule may not be true everywhere \cite{D.V.Shirkov.Theor.Math.Phys.136.1.893.2003}. Then, for a rigorous mathematical
approach, one writes the Fourier sine transform of Eq. (\ref{eq:bdtd}) as \cite{A.Erdelyi.W.Magnus.F.Oberhettinger.F.G.Tricomi.Tables.of.Integral.Transforms.McGraw.Hill.NewYork.2vols.1954}
\begin{eqnarray}\label{eq:fst}
	\alpha_s(r)=\int_0^\infty e^{-Q^2/4\kappa^2} \sin(Qr)dQ,
\end{eqnarray}

\noindent which can be solved in terms of Dawson integral function defined as \cite{H.G.Dawson.ProceedingsoftheLondonMathematicalSociety.s1.29.519.1897,M.Abramowitz.I.A.Stegun.ErrorFunctionandFresnelIntegrals.Handbook9thed.NewYork1972}
\begin{eqnarray}\label{eq:dawson}
	D(x)=e^{-x^2}\int_0^x e^{-y^2}dy^2,
\end{eqnarray}

\noindent closely related to the error function $er\!f(x)$ by \cite{M.Abramowitz.I.A.Stegun.ErrorFunctionandFresnelIntegrals.Handbook9thed.NewYork1972}
\begin{eqnarray}
	D(x)=-i\frac{\pi}{2}e^{-x^2}er\!f(ix).
\end{eqnarray}

The function defined by Eq. (\ref{eq:dawson}) can be approximated by the following series expansion depending on $x$ is close to the origin ($x<\!\!<1$) \cite{M.Abramowitz.I.A.Stegun.ErrorFunctionandFresnelIntegrals.Handbook9thed.NewYork1972,F.G.Lether.P.R.Wenston.Journal.Quantitative.Spectroscopy.Radiative.Transfer.46.4.343.1991}
\begin{eqnarray}\label{eq:dawson_series}
	D(x)=\sum_{j=0}^\infty \frac{(-1)^j2^j}{(2j+1)!!}x^{2j+1}=x-\frac{2}{3}x^3+\frac{4}{15}x^5-...,
\end{eqnarray}

\noindent where $n!!$ is the double factorial defined as
\begin{eqnarray}
	n!!=\prod_{j=0}^{k}(n-2j),
\end{eqnarray}

\noindent for $k=\lceil n/2\rceil -1$, where $\lceil x \rceil$ is the ceiling function (least integer greater than or equal to $x$). In contrast, for large $x$, the Dawson function given by Eq. (\ref{eq:dawson}) has an asymptotic behavior given by \cite{F.G.Lether.P.R.Wenston.Journal.Quantitative.Spectroscopy.Radiative.Transfer.46.4.343.1991}
\begin{eqnarray}\label{eq:dawson_asymp}
	D(x)\approx \frac{1}{2x}+\frac{1}{2^2x^3}+\frac{1\cdot 3}{2^3x^5}+\frac{1\cdot 3\cdot 5}{2^4x^7}+...
\end{eqnarray}

The main goal is to study a possible chiral symmetry restoration, implying a small value for $x=\kappa r$. For the sake of simplicity, let us suppose that $r\lesssim r_h$ fm, where $r_h$ is the hadron radius. To achieve the confinement phase of QCD, we impose the constraint $\kappa<\!\!< 1$ GeV, for which one has $\kappa r<\!\!<1$, resulting we can use the series given by Eq. (\ref{eq:dawson_series}) to represent the running coupling in $r$-space, and, in particular, $(\kappa r)^3<\!\!<1$. As shall be seen, the ratio $r/r_0$, for which $r/r_0<1$ indicates the non-confinement regime is achieved, depends strongly on the choice of $\kappa$.

It is important to notice that the mass scale is also present in the Fourier sine transform since $x=\kappa r$, which introduces the need for careful analysis to obtain the correct representation of $\alpha_s(r)$ through the Dawson function. Observe that the presence of the mass scale result, after the Fourier sine transform given by Eq. (\ref{eq:fst}), in a dimensionful running coupling written as   
\begin{eqnarray}\label{eq:alpha_r1}
	\alpha_s(r)=2\kappa \sqrt{\frac{2}{\pi}}D(\kappa r)=2\kappa\sqrt{\frac{2}{\pi}}\left[\kappa r-\frac{2}{3}(\kappa r)^3+\frac{4}{15}(\kappa r)^5-...\right].
\end{eqnarray}


Then, it is necessary to change $\bar{\alpha}_s(r)\rightarrow \alpha_s(r)/\kappa$ in the above definition to restore the dimensionless feature of the running coupling. 
Now, taking into account the stated before, one can replace $\bar{\alpha}_s(r)$ in the confinement potential given by Eq.  (\ref{eq:conf_pot})
\begin{eqnarray}
	V(r)=-\frac{4}{3}\frac{\bar{\alpha}_s(r)}{r}+\sigma r=-\frac{8}{3}\kappa\sqrt{\frac{2}{\pi}}\left[1-\frac{2}{3}(\kappa r)^2+\frac{4}{15}(\kappa r)^4-...\right]+\sigma r.
\end{eqnarray}

Considering the above constraints on $\kappa$ and $r$, one can retain the series expansion up to the second order, allowing to write the confinement potential as
\begin{eqnarray}\label{eq:pot_series}
	V(r)=-\frac{4}{3}\frac{\bar{\alpha}_s(r)}{r}+\sigma r\approx -\frac{8}{3}\kappa\sqrt{\frac{2}{\pi}}\left[1-\frac{2}{3}(\kappa r)^2\right]+\sigma r.
\end{eqnarray}

The above result indicates a confinement potential with a finite negative range given by
\begin{eqnarray}
	V(0)=-\frac{8}{3}\kappa\sqrt{\frac{2}{\pi}},
\end{eqnarray}

\noindent and dependent exclusively on the mass scale. 

\subsection{Entropy}

For the region near the minimum of the total cross section, the hadron volume is almost constant, implying the entropy can be obtained by using the Helmholtz free energy \cite{S.D.Campos.Arxiv.2020}
\begin{eqnarray}\label{eq:ent_pot}
	S_H(s)=\frac{1}{T_c}V(r),
\end{eqnarray}

\noindent where $r=r(s)$, and the transition temperature $T_c$ depends on the potential approach used and $T_c\approx 0.001$ GeV for the Cornell potential \cite{S.D.Campos.Arxiv.2020}. It is important to point out that (\ref{eq:ent_pot}) should be used only close to the minimum of $\sigma_{tot}(s)$ since the Helmholtz free energy does not necessarily vanish elsewhere. 

The main assumption here is that the entropy given by Eq. (\ref{eq:sdc_1}) can be defined through the Helmholtz free energy near the minimum of $\sigma_{tot}(s)$. Thus, one can write
\begin{eqnarray}
	S(s)=\gamma S_H(s)
\end{eqnarray}

\noindent where $\gamma$ is a real parameter. The justification of such a choice is quite difficult and is based on the fact that both definitions depend on $r$ at the minimum of $\sigma_{tot}(s)$. For the sake of simplicity, one assumes $\gamma=1$ since it acts only as a scale to the problem.

Using potential given by Eq. (\ref{eq:pot_series}) and entropy in Eq. (\ref{eq:ent_pot}), one can introduce a cut-off in the series expansion of the exponential in Eq. (\ref{eq:sdc_2}) if $V(r)/T_c<\!\!< 1$. This cut-off can be obtained for $\kappa \sim T_c<\!\!<1$ GeV. If satisfied, these constraints can indicate the unambiguous emergence of free quarks, as shall be seen. Then, one writes the following approximation for $r$
\begin{eqnarray}\label{eq:rs_final}
	r^2\approx \left\{1-\frac{1}{T_c}\left[-\frac{8}{3}\kappa\sqrt{\frac{2}{\pi}}\left(1-\frac{2}{3}(\kappa r)^2\right)+\sigma r \right]\right\} \sigma_{tot}(s),
\end{eqnarray}

\noindent according to the constraints on $\kappa$, $r$, and confinement potential adopted. Of course, different approaches for $V(r)$ and $\alpha_s(r)$ will produce distinct expressions for $r$. Moreover, there are definitions for $V(r)$ that are independent of the running coupling (see, for example, \cite{C.Quigg.J.L.Rosner.Phys.Lett.B71.153.1977}). 

\subsection{Total Cross Section}

Now, it is necessary to introduce a functional form for the total cross section. In the intermediate energy range, one can use the following simple parameterization for the total cross section \cite{S.D.Campos.Chin.Phys.C.2020}
\begin{eqnarray}\label{eq:sigma_fit}
	\sigma_{tot}(s)=a_1(s/s_0)^{a_2}+a_3\ln^{\alpha_\mathbb{P}(0)}(s/s_0),
\end{eqnarray}

\noindent where $a_1$, $a_2$, $a_3$, and $\alpha_\mathbb{P}(0)$ are free parameters. In the fitting process, one uses only the experimental data available for $pp$ total cross section above $\sqrt{s}=3.0$ GeV (including cosmic-ray data), obtained from Particle Data Group \cite{PDG-PhysRev-D98-030001-2018}. From the fitting procedure, one obtains the parameters shown in Table \ref{tab:table_1}, where $\alpha_\mathbb{P}(0)$ is typical of the hard pomeron picture \cite{V.S.Fadin.E.A.Kuraev.L.N.Lipatov.Sov.Phys.JETP44.443.1976,Y.Y.Balitsky.L.N.Lipatov.Sov.J.Nucl.Phys.28.822.1978}.

\begin{table*}[ht]
	
	{\begin{tabular}{c | c | c | c | c}
			
			\hline
			$a_1$ (mb)     ~&~ $a_2$          ~&~ $a_3$ (mb)    ~&~ $\alpha_\mathbb{P}(0)$ ~&~ $\chi^2/ndf$ \\ 
			\hline
			$52.52\pm0.38$ ~&~  $0.14\pm0.01$ ~&~ $0.91\pm0.08$~&~ $1.61\pm0.03$ ~&~ 1.77 \\
			\hline
		\end{tabular}\caption{Parameters obtained by using (\ref{eq:sigma_fit}) in the fitting procedure and assuming  $\sqrt{s_0}=1.0$ GeV.\label{tab:table_1}}}
	
\end{table*}

\begin{figure}
	\centering{\includegraphics[scale=0.5]{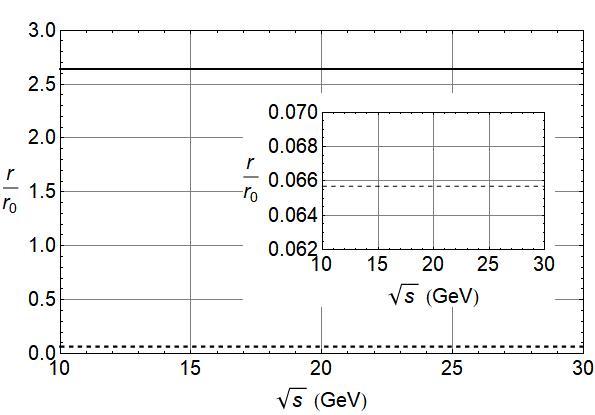}}
	\caption{Behavior of $r/r_0$ near the minimum of the total cross section. Solid line is for $\kappa =0.1$ GeV and dashed line is for $\kappa=0.002$ GeV. Inner panel shows only the curve for $\kappa=0.002$ GeV where $r/r_0\approx 0.066$, indicating the presence of free quarks in the confinement phase of QCD.}
	\label{fig:r_r0}
\end{figure}

As well-known, the hadronic total cross section decreases as the collision energy grows, from the Coulomb-Nuclear region up to some minimum located in the energy range $\sqrt{s}\approx 10\sim 30$ GeV. This behavior is attributed to the odderon exchange, also known as the odd-$C$ Regge exchange, a three-gluon state \cite{J.Bartels.Nucl.Phys.B.175.365.1980}, recently observed \cite{G.Antchev.etal.TOTEM.Coll.Eur.Phys.J.C.79.785.2019}. This leading particle exchange differentiates particle-particle from particle-antiparticle scattering and, as $\sqrt{s}$ grows, this exchange becomes less important, implying that, at sufficiently high energy, both total cross sections tend to the same value (according to the Pomeranchuk theorem). In fact, this behavior for both $\sigma_{tot}(s)$ at very high energies is attributed to the exchange of a leading particle called pomeron (not yet observed), which does not differentiate particle from antiparticle.

\subsection{Results Depending on $\kappa$}

Taking into account the approaches above performed as well as the fitting result, the only remaining free parameter of the approach is the mass scale $\kappa$.  

Figure \ref{fig:r_r0} shows the evolution of $r/r_0$ near the minimum of the total cross section according to $\sqrt{s}$. As expected, it presents an almost flat behavior since there is (almost) no variation on the hadronic total cross section in this energy range. This result indicates the distance between the $q\bar{q}$-pairs in the cell remains almost unchanged near the minimum of $\sigma_{tot}(s)$. The solid line describes the evolution of $r/r_0$ for $\kappa=0.1$ GeV (solid line) and for $\kappa=0.002$ GeV (dashed line) in the energy range $\sqrt{s}=10\sim 30$ GeV. For $\kappa=0.1$ GeV there is no emergence of a non-confinement phase since $r/r_0\approx 2.6$. In this situation, $r>1$ fm may break the approximations performed here. In contrast, for $\kappa=0.002$ GeV, the ratio $r/r_0\approx 0.066$, indicating the unequivocal emergence of chiral restoration in the confinement phase of QCD. It is important to point out that $\kappa\sim m_q$, where $m_q$ is the current quark mass (also $T_c\sim m_q/2$).

\begin{figure}
	
	\centering{\includegraphics[scale=0.31]{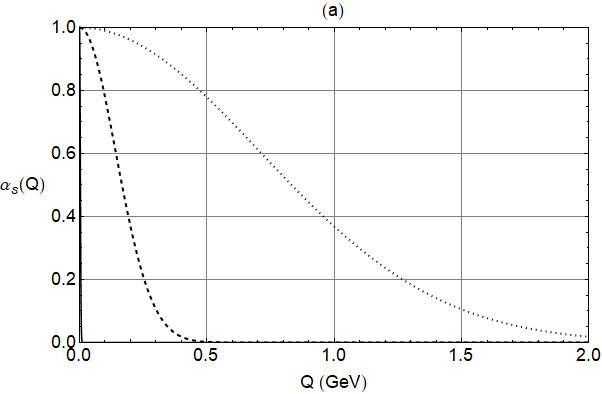}}
	\centering{\includegraphics[scale=0.31]{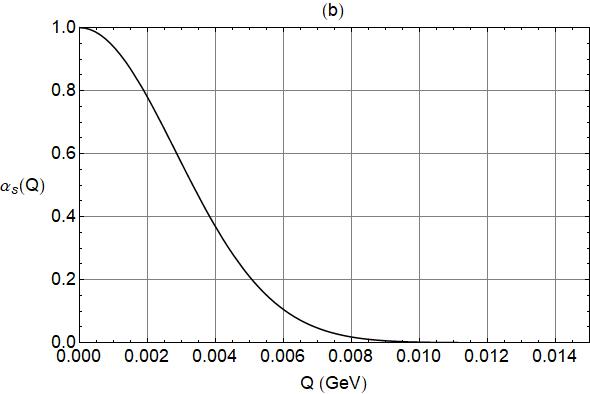}}
	
	\centering{\includegraphics[scale=0.3]{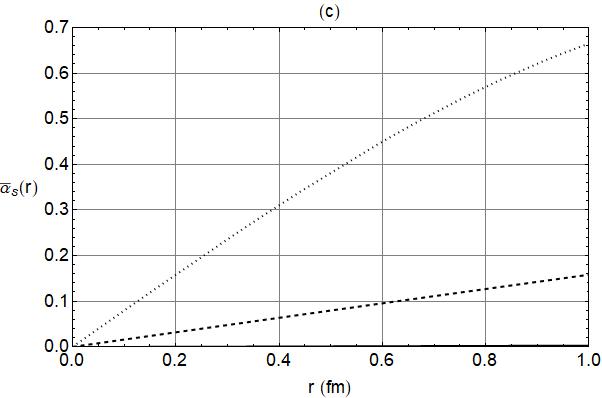}}
	\centering{\includegraphics[scale=0.31]{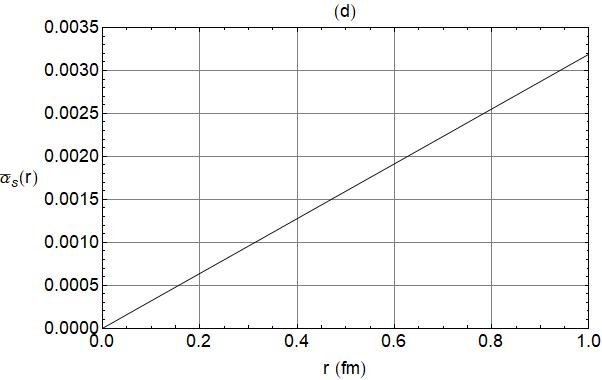}}
	
	\caption{For panels (a) and (c), solid line ($\kappa=0.002$ GeV), dashed line ($\kappa=0.1$ GeV), and dotted line ($\kappa=0.500$ GeV). Panels (b) and (d) shows the behavior of the running coupling for $\kappa=0.002$ GeV. The fast decreasing of $\alpha_s(Q)$ in panel (b) represents a very slow increase in the position space (d).}
	\label{fig:alpha}
\end{figure}

One observes that when $\kappa$ diminishes, the distance $r$ between quark and antiquark in the cell also decreases. Then, the mass scale measures the $e\!f\!f\!ective$ mass of the cell, and when $\kappa\lesssim 0.002$ GeV, the quark and antiquark are free to achieve new degrees of freedom. It does not mean that all pairs inside the hadron will achieve the non-confinement phase at this energy range: Only the pairs where $\kappa\lesssim 0.002$ GeV $\sim m_q$. The unbinding pairs are shielded by binding pairs that have not achieved the non-confinement phase, and, probably, these initial free quarks are created near the center of the hadron, explaining why we do not see these free states in the confinement regime of QCD. 

It is important to stress that the chiral restoration can also be achieved assuming, for example, $\kappa\lesssim 0.035$ GeV, resulting in $r/r_0\lesssim 0.94$. However, high values for $\kappa$ cannot ensure the validity of the truncation of the exponential series performed to obtain Eq. (\ref{eq:rs_final}). 

The appearance of free quarks in the confinement phase of QCD at the center of the colliding hadron may explain the hollowness effect \cite{dremin_1,dremin_2}, marked by the appearance of a gray area near the center of the hadron (please, see \cite{S.D.Campos.V.A.Okorokov.C.V.Moraes.Phys.Scr.95.025301.2020}, and references therein). This effect, roughly speaking, indicates that the black disk limit cannot be achieved even for asymptotic energies.

In Figure \ref{fig:alpha}, panels (a) and (b) show the behavior of $\alpha_s(Q)$ using (\ref{eq:bdtd}) and depending on $\kappa$. Solid line is for $\kappa=0.002$ GeV (almost invisible in panel (a)), dashed line is for $\kappa=0.1$ GeV, and for $\kappa=0.500$ GeV one has the dotted line. Observe that adopting $\kappa$ as a free parameter introduces the possibility of a non-confinement phase even for small $Q$. Panels (c) and (d) are for $\bar{\alpha}_s(r)$ using the same values for $\kappa$. In panel (c), the solid line for $\kappa=0.002$ GeV is almost flat compared to the other curves, while in panel (a), considering the same $\kappa$, the solid line presents a fast decrease. 




\section{\label{fr}Final Remarks}

In the BKT phase transition, the vortex-antivortex pair are unbinding at the critical temperature. An analogy for this phase transition can be performed, treating $q\bar{q}$-pairs as the vortex-antivortex pairs \cite{S.D.Campos.Arxiv.2020}. As obtained in this work, under some physical conditions, the $q\bar{q}$-pairs can no longer exist even in the confinement phase of QCD. The possible chiral symmetry restoration in the confinement phase should be understood as the co-existence of free quarks and colorless states during the collision. 

It is interesting to point out that a few years ago was predicted both the existence of quarkyonic matter (which is not a quark) and the possibility that it suffers a chiral symmetry restoration in the confinement phase of QCD \cite{L.McLerran.R.D.Pisarski.Nucl.Phys.A796.83.2007,L.YA.Glozman.R.F.Wagenbrunn.Mod.Phys.Lett.A23.2385.2008}. According to McLerran and Pisarski \cite{L.McLerran.R.D.Pisarski.Nucl.Phys.A796.83.2007}, for a large number of colors $N$, it is reasonable that for a large chemical potential $\mu_q$ and low temperature, a confining regime exists as well as the preservation of chiral symmetry.

Using the running coupling constant defined in the light-front approach for QCD, we obtain a confinement potential depending on the mass scale $\kappa$, considered here as the effective mass of the $q\bar{q}$-pair. The Helmholtz free energy near the minimum of the total cross section allows connecting entropy with the confinement potential. A fitting procedure is implemented to $\sigma_{tot}(s)$ using $pp$ experimental data. Then, the ratio $r/r_0$ is analyzed depending only on $\kappa$. 

Considering the mass scale as a free parameter, one assumes that $\kappa$ represents the effective mass contained in the cell. For $\kappa=0.002$ GeV, the ratio $r/r_0\approx 0.066$ indicates the emergence of chiral symmetry restoration in the confinement phase of QCD. It is important to stress that $\kappa$ is close to the current quark mass, i.e. at the chiral symmetry restoration, one may expect $\kappa\sim m_q$. On the other hand, for $\kappa\gtrsim  0.035$ GeV, there is no emergence of free quarks in the confinement phase of QCD. Notice, furthermore, that in the range $0.002$ GeV$\lesssim\kappa\lesssim 0.035$ GeV the chiral symmetry restoration is allowed if the approximations performed could be ensured.

As a final conjecture, the decreasing of $\kappa$, resulting in the decreasing of $r$, may be a consequence of the odderon exchange occurring from the Coulomb-Nuclear region up to the minimum of $\sigma_{tot}(s)$. As the collision energy smoothly grows above the minimum of $\sigma_{tot}(s)$, the odderon is no longer the leading particle exchange, which could imply a decrease in the emergence of free quarks. However, above the minimum of $\sigma_{tot}(s)$ the creation of free quark through odderon exchange, although less effective due to the diminishing (not the vanishing) in the odderon exchange, possibly continues up to the hadron dissociation (at the saturation scale). Since the emergence of free quarks leads to the increasing of entropy inside the hadron, then the growth of $\sigma_{tot}(s)$ above the minimum may be caused (in thermodynamics language) by the increase of entropy, which leads to the hadron dissociation at the saturation scale \cite{S.D.Campos.A.M.Amarante.In.J.Mod.Phys.A35.2050095.2020}.

\section*{Acknowledge}

SDC thanks to UFSCar for the financial support.

\end{document}